\documentclass[fleqn,usenatbib]{mnras}

\usepackage{newtxtext,newtxmath}

\usepackage[T1]{fontenc}

\DeclareRobustCommand{\VAN}[3]{#2}
\let\VANthebibliography\thebibliography
\def\thebibliography{\DeclareRobustCommand{\VAN}[3]{##3}\VANthebibliography}

\usepackage{graphicx}	
\usepackage{amsmath} 

\usepackage{amssymb} 

\usepackage{ulem}       
\usepackage{xcolor}     
\usepackage{subcaption} 

\newcommand{\HI}{H{\sc i}}	
\newcommand{\kms}{\,km\,s$^{-1}$}
\newcommand{\Msol}{\rm{M}_{\odot}}
\newcommand{\Htwo}{H$_{2}$}

\defcitealias{Obreschkow2016}{O16}
\defcitealias{Hardwick2022}{Paper 1}

\title[The connection between j, M and gas fraction]{xGASS: The connection between angular momentum, mass and atomic gas fraction in nearby galaxies}

\author[J. A. Hardwick et al.]{Jennifer A. Hardwick,$^{1,2}$\thanks{E-mail: \href{mailto:jennifer.hardwick@icrar.org}{jennifer.hardwick@icrar.org}}
Luca Cortese,$^{1,2}$
Danail Obreschkow$^{1}$
and Barbara Catinella$^{1,2}$
\\
$^{1}$International Centre for Radio Astronomy Research (ICRAR), University of Western Australia, Crawley, WA 6009, Australia\\
$^{2}$Australian Research Council, Centre of Excellence for All Sky Astrophysics in 3 Dimensions (ASTRO 3D), Australia\\
}

\date{Accepted 2022 August 26. Received 2022 August 17; in original form 2022 June 22}

\pubyear{2022}

\begin{document}
\label{firstpage}
\pagerange{\pageref{firstpage}--\pageref{lastpage}}
\maketitle

\begin{abstract}
We use a sample of 559 disc galaxies extracted from the eXtended GALEX Arecibo SDSS Survey (xGASS) to study the connection between baryonic angular momentum, mass and atomic gas fraction in the local Universe.
Baryonic angular momenta are determined by combining \HI\ and \Htwo\ integrated profiles with two-dimensional stellar mass surface density profiles. 
In line with previous work, we confirm that specific angular momentum and atomic gas fraction are tightly correlated, but we find a larger scatter than previously observed.
This is most likely due to the wider range of galaxy properties covered by our sample. 
We compare our findings with the predictions of the analytical stability model developed by Obreschkow et al. and find that, while the model provides a very good first-order approximation for the connection between baryonic angular momentum, mass and gas fraction, it does not fully match our data. 
Specifically, we find that at fixed baryonic mass, the dependence of specific angular momentum on gas fraction is significantly weaker, and at fixed gas fraction, the slope of the angular momentum vs. mass relation is shallower than what was predicted by the model. 
The reasons behind this tension remain unclear, but we speculate that multiple factors may simultaneously play a role, all related to the fact that the model is not able to encapsulate the full diversity of galaxy properties in our sample.

\end{abstract}

\begin{keywords}
galaxies: disc -- galaxies: evolution -- galaxies: ISM -- galaxies: kinematics and dynamics
\end{keywords}



\section{Introduction} \label{section: intro}

Understanding how galaxies grow in mass and structure across cosmic time, and if/how these processes can regulate the gas-star formation cycle, is a key step towards understanding how galaxies form and evolve.
At least for disc galaxies, in our current theoretical framework, the growth of angular momentum (AM) of discs is tightly connected to the ability of galaxies to accrete gas and use it for star formation \citep[e.g.,][]{Mo1998, Boissier2000}. 
As such, we expect some kind of correlation between the AM of discs and their cold gas fractions. 
While this link was initially suggested by early analytical studies, now it has been shown in both cosmological simulations and observations. For example, several independent studies have now shown that,  at fixed mass, the spread in AM correlates with the amount of atomic gas available \citep[e.g.,][]{Huang2012,Obreschkow2016,Lagos2017, Lutz2018, Stevens2018, Wang2019,Murugeshan2019,Pina2021,Kurapati2021,Hardwick2022}.

The next step is clearly to move beyond simple correlations and understand whether or not they trace any underlying physical connection between AM and gas content.
This task is far from being trivial. 
From an observational point of view, given that most of the observed galaxy properties are interconnected, identifying the potential physical drivers of scaling relations demands a very large number statistics. 
In fact, only in this way can we look for trends after controlling for various galaxy properties (e.g., mass, morphology and environment). 
Unfortunately, as discussed below, we are still lacking information for both AM and gas content for a large enough sample to properly dissect the AM-mass-gas fraction plane, in the same way that has been done for other well-known scaling relations \citep[][]{Bernardi2003, Courteau2007,Saintonge2022}.
This is equally difficult from a theoretical point of view, as there is limited ability for cosmological simulations to trace multiple and interconnected physical processes, which makes it challenging to identify the dominant driver of scaling relations.
This is where simpler analytical models can provide a complementary view to the problem, as they allow us to test whether or not a simple framework of galaxy evolution is able to reproduce our observed trends.

Intriguingly, in the last few years, the analytical approach has been extremely popular in our quest for understanding the potential physical link between AM and gas content. 
In particular, \citet[][hereafter \citetalias{Obreschkow2016}]{Obreschkow2016} have put forward an analytical model showing how this correlation can be interpreted from a disc stability point of view (see also \citealp{Romeo2020} for an alternative approach).
They relate the baryonic AM and mass of a galaxy to its atomic gas content by using the \cite{Toomre1964} stability parameter. 
Specifically, they assume galaxies to be thin axisymmetric exponential discs, and calculate regions where gas is stable to collapse and could remain atomic, based on the galaxies baryonic AM. 
The total atomic gas fraction of a system is then obtained by integrating the gas mass in these stable regions.
Many previous works have shown that the \citetalias{Obreschkow2016} model matches \HI-selected disc galaxies in observations \citep{Lutz2018, Murugeshan2019, Dzudzar2019, Murugeshan2020, Li2020, Murugeshan2021, Kurapati2021}, and is qualitatively consistent with predictions from cosmological simulations \citep{Wang2018, Stevens2018, Stevens2019}. 
This has been used as a strong support to the idea that AM is connected to the amount of atomic gas fraction in galaxies.

While promising, the agreement between observations and the \citetalias{Obreschkow2016} model has so far been established primarily for pure discs, gas-rich galaxies, which may not be representative of the local galaxy population, particularly at stellar masses higher than 10$^{10}$ M$_{\odot}$ (e.g., \citealp{Catinella2018, Cook2019}). 
As such, it is important to extend previous analyses to samples spanning significantly larger ranges of morphologies and gas fractions. 

In fact, recent work on the topic has started providing us with hints of a potential tension between observations and the \citetalias{Obreschkow2016} model.
\cite{Pina2021,Pina2021b} used a sample of 157 galaxies with resolved \HI\ information to investigate the link between the scatter of the AM-mass relation (i.e., the \citealp{Fall1983} relation) and \HI\ gas fraction. 
While they find a significant correlation between the two quantities, they were not able to match their finding with the \citetalias{Obreschkow2016} model. 
In \citet[][hereafter referred to as \citetalias{Hardwick2022}]{Hardwick2022}
we investigated the stellar AM - mass relation for the largest representative sample of nearby galaxies with \HI\ measurements to date (the extended GALEX Arecibo SDSS survey;  \citealt{Catinella2018}). 
Again, we found a strong correlation between \HI\ gas fraction and stellar specific AM, but we show how this correlation becomes weaker for high mass galaxies with a photometric bulge component, potentially highlighting some deviations from the expectations for pure disc systems. 
However, to test these conclusions, it is critical to perform a careful comparison with the \citetalias{Obreschkow2016} model that not only focuses on the stellar specific AM, but also considers the full baryonic AM of galaxies.

Thus, in this paper, we build on the work done in \citetalias{Hardwick2022} and investigate in detail whether or not the \citetalias{Obreschkow2016} model can provide a quantitatively accurate representation of the correlation between AM, mass and gas fraction, over a large range of stellar masses and gas content.

This paper is set out as follows.
In Section \ref{section: sample} we outline the sample used and summarise how these measurements were determined.
In Section \ref{section: methods} we describe how we calculate baryonic mass and specific AM needed to investigate the \citetalias{Obreschkow2016} relation.
Section \ref{section: results} gives the key results of this work on the \citetalias{Obreschkow2016} relation and our empirical fit before discussing their implications in Section \ref{section: discussion}.
We conclude and summarise our findings 
in Section \ref{section: summary}.

\section{Sample}
\label{section: sample}

The extended GALEX Arecibo SDSS Survey (xGASS; \citealt{Catinella2010, Catinella2018}) has the most sensitive \HI\ observations for a large representative sample in the local Universe, covering stellar masses between $10^{9}\, \Msol$ and $10^{11.5}\, \Msol$ across the redshift range $0.01 < z < 0.05$.
xGASS was selected from the Sloan Digital Sky Survey, Data Release 6 (SDSS DR6; \citealt{Adelman-McCarthy2008}) spectroscopic catalogue overlapping with the GALaxy Evolution eXplorer (GALEX; \citealt{Martin2005}) sky footprint. 
\HI\ masses and velocity widths were determined by taking observations with the 305m Arecibo single-dish telescope. 
Each observation was taken until \HI\ was detected or a gas fraction limit of 2\% to 10\% (depending on stellar mass) was reached. 

In this work, we use the sub-sample of galaxies extracted from the xGASS survey that were detected in \HI\ (not including any with possible \HI\ confusion) and have inclinations greater than 30 degrees, for which stellar AM estimates have been determined in \citetalias{Hardwick2022}. 
We refer the reader to this paper for details on the technique and here provide just a brief summary.  

We calculated $j_{\star}$ using spatially unresolved \HI\ observations and reconstructed surface density profiles integrated out to 10 effective radii ($R_{e}$).
Rotation curves were assumed to be flat with maximum rotational velocities equal to half the \HI\ width (corrected for inclination) measured at 50\% of the peak flux.
In this paper, we focus on the AM of the disc only and restrict our sample to galaxies with a disc component as identified by 2D decompositions \citep{Cook2019}: 559 galaxies in total.

Molecular gas masses are taken from xCOLD GASS (extended CO Legacy Database for GASS;
\citealt{Saintonge2011,Saintonge2017}).
This database was created as a follow up of xGASS to determine molecular gas masses.
CO (1-0) observations were taken on the IRAM 30m telescope for 45\% of the xGASS galaxies (these were randomly selected).
In our sub-sample, we have 248 galaxies that overlap with the xCOLD GASS data.
Of these overlapping galaxies, 196 galaxies have a CO (1-0) detection which can be used to determine a molecular gas mass using a metallicity-dependent conversion function \citep{Accurso2017}, while the remaining 52 galaxies have upper limits ($3\sigma$).
For galaxies that were not observed with xCOLD GASS, or only had upper limits, we used estimated molecular gas masses from star formation rates (SFRs) in our analysis (see Section \ref{section: baryonic_mass}).

Global SFRs and galaxy environment measures are taken from the analysis of \cite{Janowiecki2017}. 
They determined SFRs by combining GALEX near-ultraviolet and WISE \citep{Wright2010} mid-infrared photometry and galaxy environments from the \cite{Yang2007} group catalogues of SDSS.
For reference, the star-forming main sequence (MS) for xGASS is presented in \cite{Catinella2018} and \cite{Janowiecki2020}.
To calculate the offset from the MS ($\Delta$MS) for each galaxy, we take the difference between the MS (as is defined in \citealt{Catinella2018}) and the specific SFR (SFR$/M_{\star}$) of the galaxy, at its stellar mass.

\section{Methods}
\label{section: methods}

While in \citetalias{Hardwick2022} we focus on stellar AM, in order to fully investigate the physical connection between gas content and disc stability we need to quantify the total baryonic mass and baryonic AM of galaxies. 
Therefore, in this section, we describe how these quantities are derived for our sample.

\subsection{Baryonic Mass}
\label{section: baryonic_mass}

We assume that the baryonic component of disc galaxies can be fully represented by a combination of stars, atomic gas and molecular gas.
We use molecular gas masses from xCOLD GASS, (this includes a helium contribution). 
Atomic gas is comprised of atomic hydrogen (\HI, taken from xGASS) and helium (assumed to be 35\% of the \HI).
We also assume that all the hydrogen and inferred helium are located in the disc. 
Using bulge-to-disc decompositions, the stellar mass of the disc component can be separated (for details see \citealt{Cook2019} and \citealt{Hardwick2022}).
Therefore, the baryonic mass of the disc is given as follows:
\begin{equation}
        M_{\rm{bar,D}} = M_{\star,D} + M_{\rm{mol}} + 1.35M_{HI}
        \label{eq: Mbar}
\end{equation}
where $M_{\star,D}$ is the stellar mass of the disc, $M_{HI}$ is the mass of atomic hydrogen gas and $M_{\rm{mol}}$ is the mass of molecular gas.

\begin{figure}
    \centering              
    \includegraphics[width=0.5\textwidth]{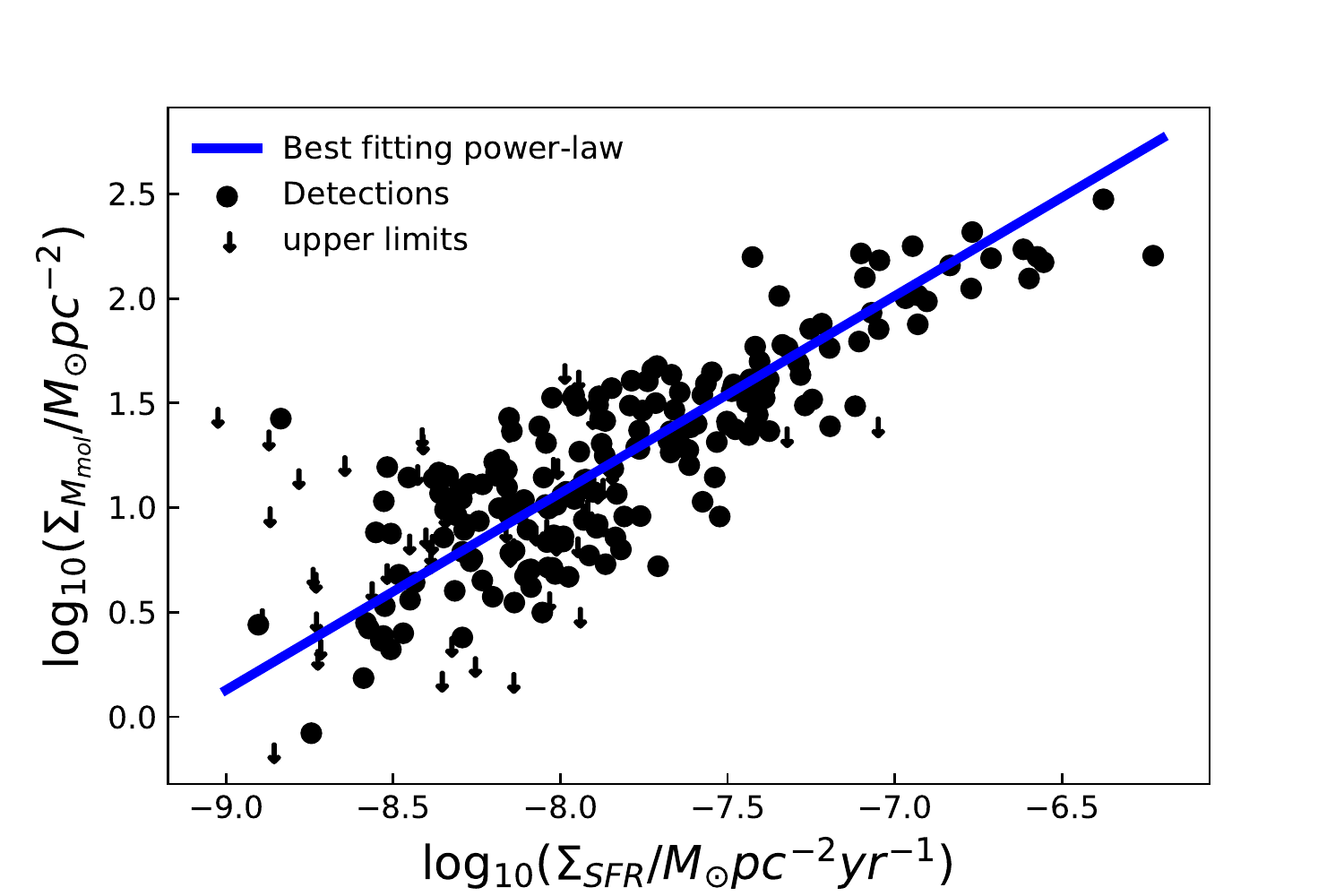}
    \caption{The inverted Kennicutt-Schmidt relation for the 251 galaxies in our sample with observations available from xCOLD GASS. Points are detections with downward arrows indicating non-detections at their 3$\sigma$ upper limit. The best-fit  power-law to the detections is shown as a thick blue line, with its formula given in Equation \ref{eq: Kennicutt_Schmidt}}
    \label{fig: Kennicutt_Schmidt}
\end{figure}

Despite all our galaxies having \HI\ and stellar masses, only 45\% of our sample has molecular gas masses from xCOLD GASS. 
Therefore, the remaining 363 galaxies in our sample need their molecular gas mass estimated.
We use the SFR surface density ($\Sigma_{\rm{SFR}} = \rm{SFR} / \pi R_{e}^2$) vs. molecular mass surface density ($\Sigma_{\rm{mol}} = M_{\rm{mol}} / \pi R_{e}^2$) relation (i.e., the inverted Kennicutt-Schmidt law, \citealt[][]{Kennicutt1989, Schmidt1959}) to estimate molecular gas for the galaxies not observed or with only upper limits in xCOLD GASS.
Here, $R_{e}$ is the r-band half-light radius of the total galaxy.
Figure \ref{fig: Kennicutt_Schmidt} shows this relation for the subset of our sample observed with xCOLD GASS.
For this sub-sample, the SFR and molecular gas surface densities are related by the following formula;
\begin{equation}
    \log_{10}(\Sigma_{M_{\rm{mol}}}/ \Msol \rm{pc}^{-2}) = 0.94 \log_{10}(\Sigma_{\rm{SFR}} / \Msol \rm{pc}^{-2} yr^{-1}) +8.61
    \label{eq: Kennicutt_Schmidt}
\end{equation}
which was determined by fitting a power-law (blue line) in \textsc{hyper-fit}\footnote{This same code has been used to determine all the best fits in this paper.} \citep{Robotham2015} to the galaxies with \Htwo\ detections.
Using equation \ref{eq: Kennicutt_Schmidt}, we obtain molecular gas masses from SFRs for the galaxies where a CO detection is not available.
For two of the galaxies in our sample, we do not have SFRs or \Htwo\ detections (GASS123007 and GASS38198). 
For these, we determine the molecular gas mass by assuming that their \HI-to-\Htwo\ mass ratio is the average value observed in our sample (i.e., 20\%).
As \Htwo\ contributes very little to both $M_{\rm{bar}}$ and $j_{\rm{bar}}$, these assumptions do not affect our results.

\subsection{Specific Angular Momentum of the Baryons}
\label{section: baryonic_AM}
    
To calculate the baryonic specific AM of the disc ($j_{\rm{bar,D}}$) we take the mass-weighted average of the specific AM of each component; stars, \HI\ and molecular gas, as follows:
\begin{equation}
    j_{\rm{bar,D}} = \frac{j_{\star,D} M_{\star,D} + j_{\rm{mol}} M_{\rm{mol}} + 1.35 j_{HI} M_{HI}}{M_{\star,D}  + M_{\rm{mol}} + 1.35M_{HI}},
    \label{eq: jbar_expanded}
\end{equation}   
where $j_{\star,D}$, $j_{\rm{mol}}$ and $j_{HI}$ are the specific AM of the disc stars, molecular gas and \HI, respectively.    
We use the values of $j_{\star,D}$ that are available from \citetalias{Hardwick2022}.

We assume that $j_{\rm{mol}}$ is equal to $j_{\star,D}$.
This is because it has been shown in works such as \cite{Bigiel2012} and \cite{Smith2016} that the surface density profile of the \Htwo\ follows closely that of the stars. 
The specific AM is primarily set by the surface mass density distribution (assuming that the rotation curves are similar for both components), so it follows that the specific AM of the \Htwo\ is the same as the stars.    

To determine $j_{HI}$ from spatially unresolved data, we proceed as follows.
Firstly, we assume the \HI\ surface density profile ($\Sigma_{HI}$) to be universal for all galaxies. 
This was shown to be a good approximation for the sample in \cite{Wang2016} when radii were normalised by the \HI\ diameter.
To determine the \HI\ diameter we use the intrinsically tight \HI\ mass to size relation from \cite{Wang2016}:
\begin{equation}
    \log_{10}(D_{HI}/ \rm{kpc}) = (0.506 \pm 0.003) \log_{10}(M_{HI}/ \rm{M}_{\odot}) - (3.293 \pm 0.009),
    \label{eq: WangHIMassSize}
\end{equation}
where $D_{HI}$ is the diameter of the \HI\ disc measured at a surface density of $\Sigma_{HI} = 1\, \Msol\, pc^{-2}$.
We tested the reliability of this approach by also allowing for a variation in diameter consistent with 1$\sigma$ above and below the relation.
Accounting for the scatter in the \HI\ mass-diameter relation changes $j_{HI}$ by a mean difference of $\pm 0.06$ dex.
The $\Sigma_{HI}$ profiles from \cite{Wang2016} are only given out to $1.3 R_{HI}$, (where $R_{HI} = 0.5 D_{HI}$), which for the majority of our galaxies (92\% of the sample) is less than the extent used to estimate $j_{\star}$ (i.e., $10 R_{e}$).
For consistency with all other quantities $j_{HI}$ is also calculated within $10R_{e}$.
To do so, we linearly extrapolate the universal $\Sigma_{HI}$ profile (in log-log space). This extrapolation slightly increases the value of $j_{HI}$ (less than 0.03 dex).
Conversely, for the 46 galaxies, where $10 R_{e} < 1.3 R_{HI}$, $j_{HI}$ is reduced by an average of 0.02 dex up to a maximum of 0.15 dex, when compared to $j_{HI}$ calculated within $1.3R_{HI}$.

Lastly, we assume that the velocity profile is constant for all radii and given by the width of the \HI\ emission line. 
This is the same assumption made in \citetalias{Hardwick2022} for calculating stellar specific AM.
This assumption was tested in \citetalias{Hardwick2022} by comparing constant velocity profiles to the template rotation curves in \cite{Catinella2006}, where we showed that the most extreme case of low-mass dwarf galaxies (i.e. $M_{\star} \approx 10^{9}M_{\odot}$), had a maximum systematic offset in j of $\sim 0.08$ dex, while high-mass galaxies had less discrepancy ($< 0.04$ dex). 
Therefore, this assumption only introduces minor systematic effects which do not impact our result (for more details see \citetalias{Hardwick2022}).
We emphasise that, despite using a constant velocity profile, our approach does not rely on the approximation of $j \propto V_{max} R_{e}$ \citep{Romanowsky2012}, and instead estimates j by directly integrating the AM profile obtained from the surface density profiles of each baryonic component. 
We found this to be a more accurate way to determine AM, as the \cite{Romanowsky2012} approximation can overestimate a galaxies AM by up to 25\% at low masses \citep[e.g.,][]{Obreschkow2014, Bouche2021}.

Once all of these assumptions are combined, we obtain $j_{HI}$.
When comparing $j_{\star, D}$ and $j_{HI}$, on average $j_{HI}$ is roughly twice the value of $j_{\star,D}$, although there is a considerable spread ($\sigma = 0.22$ dex).
Note that this factor $\sim$2 was also found in \textsc{things} \citep{Obreschkow2014} using resolved \HI\ kinematic maps.

We reiterate that although we have outlined many assumptions in this subsection, these choices are not likely to affect our results.
Although we assume that stars and gas share the same rotation curve (by using \HI\ line widths to calculate AM for both), this does not impact our results. Stellar rotation curves rise more slowly than cold gas ones due to asymmetric drift, so our estimates of $j_{\star,D}$ may be slightly overestimated. However, $j_{\rm{bar}}$ would not change considerably if $j_{\star,D}$ was altered slightly.
In addition, assumptions made when calculating $j_{\star,D}$ were robustly tested in \citetalias{Hardwick2022}.
As mentioned above, the assumptions involving $j_{HI}$ have been tested, and even if these assumptions are altered, they do not affect the result significantly. 
In addition, the molecular mass is only a small component of our galaxies, such that excluding molecular material, only alters $M_{\rm{bar}}$ by 0.02 dex and $j_{\rm{bar}}$ by 0.01 dex on average. 
In summary, the typical uncertainty on our estimate of $j_{\rm{bar}}$ is 0.14 dex, which is dominated by the assumptions made on $j_{HI}$.
This is comparable to the statistical error on $M_{\rm{bar}}$, i.e., 0.11 dex.

\section{Results}
\label{section: results}

In this section, we investigate the connection between $j_{\rm{bar}}$, $M_{\rm{bar}}$ and atomic gas fractions.

\subsection{Baryonic Fall Relation}

\begin{figure*}
    \centering
    \includegraphics[width=\textwidth]{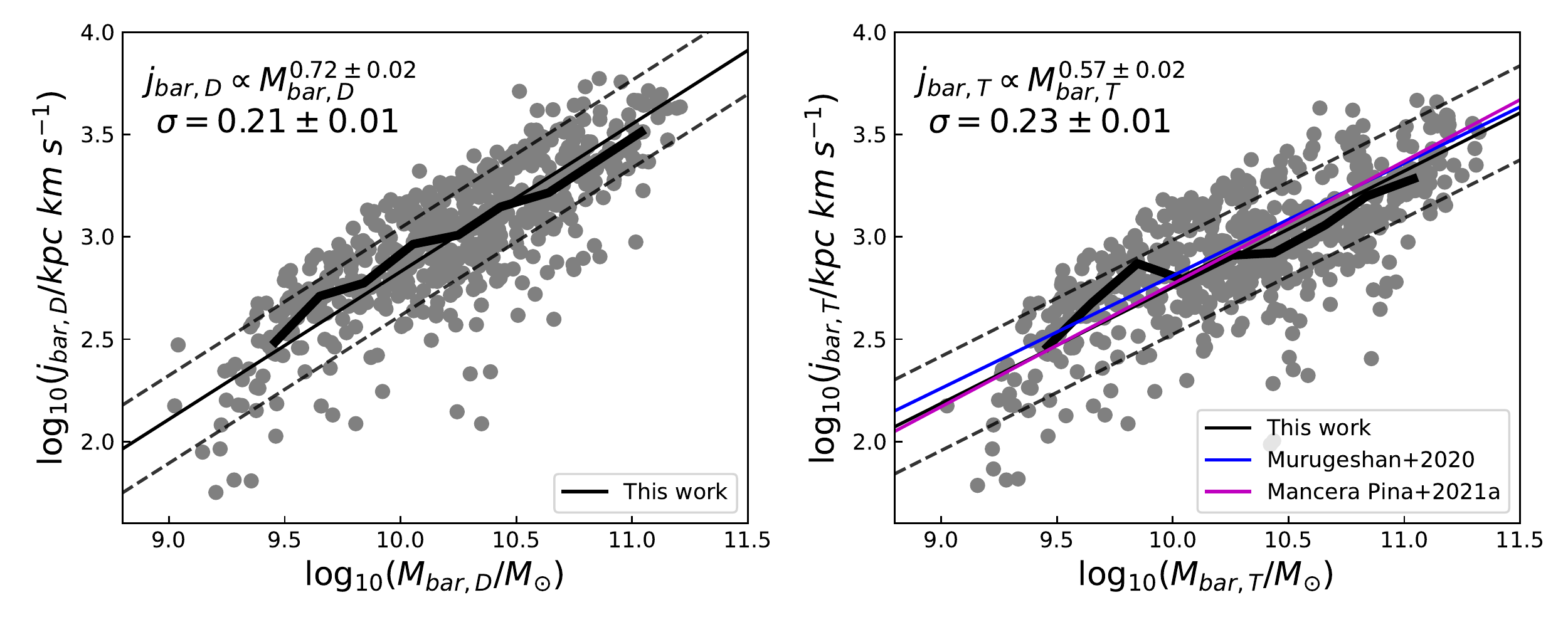}
    \caption{The baryonic Fall relation of the disc component (left) and the whole galaxy (i.e., disc plus bulge component; right). The best-fitting linear relation is shown as a thin black line, with  the 1$\sigma$ vertical scatter indicated by the dashed lines. The median in bins of baryonic mass (width = 0.2 dex) is shown as a thick black line (bins containing less than 10 galaxies are not shown). In the right panel, we compare to the work of \citet{Murugeshan2020} in blue and \citet{Pina2021} in magenta.}
    \label{fig: BaryonicFall}
\end{figure*}

As a first step, in Figure \ref{fig: BaryonicFall} we show the baryonic Fall relation for galaxies in xGASS.
The left panel shows the baryonic mass and specific AM for just the disc component ($M_{bar,D}$ and $j_{bar,D}$), while on the right we show the total (i.e., disc and bulge component; $M_{bar,T}$ and $j_{bar,T}$).
Although the majority of the paper will focus on just the disc components, we also show the right panel to compare with previous works.
The best-fit linear relation to our data is shown as a thin black line, with 1$\sigma$ scatter shown by the dashed lines.
The blue line shows the baryonic Fall relation found by \cite{Murugeshan2020} and the magenta line is from \cite{Pina2021}, which both agree well with the best-fitting relation to our work. 
This is not a trivial result given the difference in samples and techniques used to determine this relation. 
Therefore, the agreement with previous work gives us confidence that our approach is solid and we can move forward in our analysis.

As for previous works, we suggest caution when interpreting the slopes of the baryonic Fall relation at face value, as our sample is not selected by baryonic mass.
xGASS is selected to only include galaxies with stellar masses greater than $10^9\, \Msol$ with an oversampling at the high stellar mass end to increase statistics (as shown in \citetalias{Hardwick2022} this distribution is conserved for our sub-sample).
As such, the distribution in baryonic mass of our sample is not representative, with a deficiency of galaxies at lower baryonic masses.
Below a baryonic mass of approximately $10^{9.8}\, \Msol$, at fixed baryonic mass, our sample will preferentially miss galaxies at high gas fractions, whose stellar mass lies below the cut of $10^{9}\, \Msol$. 
In addition, we know from the results of works such as \citet[][see also \citetalias{Hardwick2022}]{Pina2021} that galaxies with higher gas fractions also have higher $j_{\rm{bar}}$.
Therefore, we would expect the Fall relation to be shallower for a sample that is uniformly distributed in baryonic mass.
This echoes the conclusions of \citetalias{Hardwick2022}, which found that the sample used to determine a Fall relation will affect the slope obtained.

\subsection{\HI\ content predicted from local stability} \label{section: fatm-q}

\citetalias{Obreschkow2016} developed a formalism to look at the connection between specific baryonic AM, baryonic mass and atomic gas. 
This resulted in a dimensionless effective stability parameter q;
\begin{equation}
    q \equiv \frac{j_{\rm{bar,D}} \sigma_{HI}}{G M_{\rm{bar,D}}},
    \label{eq: q}
\end{equation}
where $j_{\rm{bar,D}}$ is the baryonic specific AM of the disc component, $\sigma_{HI}$ is the one-dimensional velocity dispersion of \HI, G is the gravitational constant and $M_{\rm{bar,D}}$ is the baryonic mass of the disc component.
We use $M_{\rm{bar,D}}$ and $j_{\rm{bar,D}}$ as defined in sections \ref{section: baryonic_mass} and \ref{section: baryonic_AM} respectively. 
$\sigma_{HI}$ is assumed to be a constant 10 \kms, which is set by the temperature of the warm neutral medium (following the assumption of \citetalias{Obreschkow2016}).
For a flat exponential disc, the q parameter relates to the expected fraction of atomic gas via;
\begin{equation}
    f_{\rm{atm}} = \rm{min}\{ 1, 2.5 \ q^{1.12} \}.
    \label{eq: O16Relation_fatm-q}
\end{equation}
Here the fraction of atomic gas is defined as the fraction of disc baryonic mass in neutral atomic hydrogen and helium;
\begin{equation}
    f_{\rm{atm}} = \frac{1.35 M_{HI}}{M_{\rm{bar,D}}}.
    \label{eq: fatm}
\end{equation}
This stability model assumes that any atomic gas that is unstable (following the \citealt{Toomre1964} stability parameter Q) will collapse to form stars.
As a consequence, if $f_{\rm{atm}} > \rm{min}\{ 1, 2.5 \ q^{1.12} \}$, then the galaxy is over-saturated in \HI\ and is bound to form stars quickly (i.e., within a giant molecular clouds free-fall time).
However, in the opposite case ($f_{\rm{atm}} < \rm{min}\{ 1, 2.5 \ q^{1.12} \}$) a galaxy will remain stable, with no self-regulating processes to alter its atomic gas fraction.
Therefore, the \citetalias{Obreschkow2016} relation (equation \ref{eq: O16Relation_fatm-q}) can be thought of as an upper limit of the amount of atomic gas that a galaxy can hold (for further details of the models assumptions see \citetalias{Obreschkow2016}).

\begin{figure}
    \centering
    \includegraphics[width=0.5\textwidth]{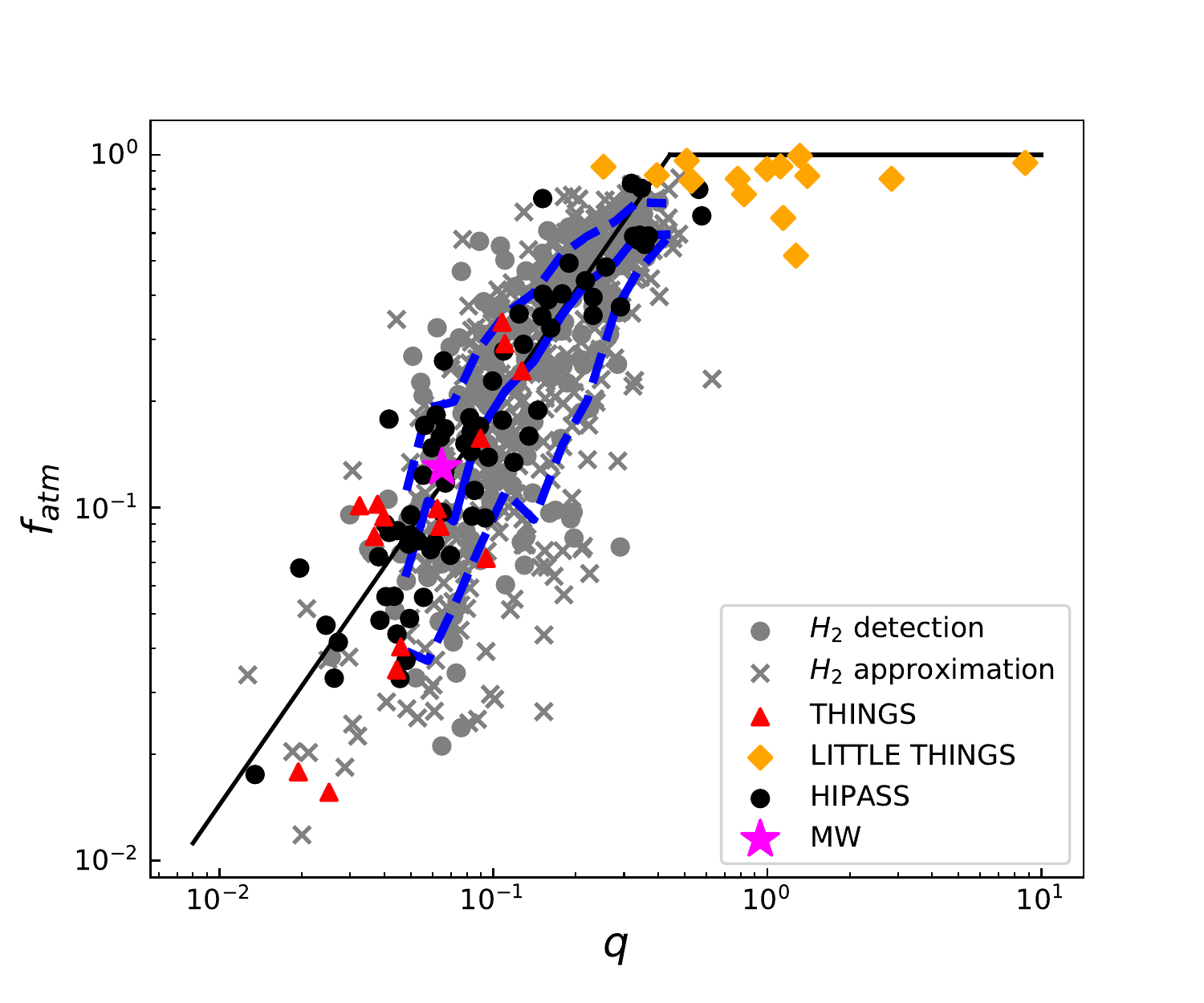}
    \caption{The distribution of our xGASS sample in the $f_{atm}$-$q$ plane (grey symbols). Circles show galaxies with \Htwo\ detections, while crosses are objects for which \Htwo\ masses have been estimated from SFRs as discussed in Section \ref{section: methods}. The blue lines show the median (solid) and 16th and 84th percentiles (dashed) for our sample, while the black line shows the theoretical relationship between q and $f_{atm}$ given in Equation \ref{eq: O16Relation_fatm-q}. The other symbols indicate the samples used in \citetalias{Obreschkow2016} for their comparison with the model.}
    \label{fig: O16_litcompare}
\end{figure}
Figure \ref{fig: O16_litcompare} shows the relationship between the q stability parameter and $f_{atm}$.
We compare our sample (grey) to the data presented in \citetalias{Obreschkow2016} (coloured points) and their model (black line).
The binned median of our data (thick blue line) shows good agreement with the theoretical relation shown in black, however, our data show a larger scatter.
For our data the median 16th to 84th percentile is 0.52 dex, whereas it is 0.27 dex for the original \citetalias{Obreschkow2016} data.
This result does not depend on the method used to determine $M_{\rm{mol}}$, as $f_{atm}$ only weakly depends on the molecular gas mass, and is mostly constrained by the \HI\ mass. 
The significant scatter observed for xGASS is an important result, as previous work have used the small observed scatter in the $f_{atm}-q$ relation to support a scenario in which $q$ is the primary galaxy property physically connected to gas fraction \citep[e.g.,][]{Obreschkow2016,Lutz2018,Murugeshan2019}. 
Instead, for our sample, the $f_{atm}-q$ relation shows a spread comparable to that of the well-known relation between gas fraction and $NUV-r$ colour (median 16th to 84th percentile 0.49 dex).

As mentioned above, the \citetalias{Obreschkow2016} model predicts that galaxies will not lie above the relation for long, as their \HI\ gas will be unstable and collapse to form stars. 
\citetalias{Obreschkow2016} expect most galaxies to lie within 40\% of the analytical relation given the approximate uncertainties of the model, with only 6 out of their 105 galaxies lying above this point.
In our data, above this limit, we have roughly double the fraction (61/559) than found in \citetalias{Obreschkow2016}.
To understand if this is simply due to issues with the data, we analysed the 61 outlier galaxies via the Tully-Fisher relation \citep[][]{Tully1977} and found that only 11 of them appear to have underestimated rotational velocities.
Therefore, as this issue does not affect the rest of the population scattering above the theoretical line, it suggests that the larger scatter compared to what is seen in \citetalias{Obreschkow2016} is not due to data quality but intrinsic to our sample.

Previous works have shown that the down-ward scatter of the $f_{atm} - q$ relation can be due to environmental effects removing the gas from the disc without affecting the stability parameter \citep[e.g.][]{Li2020, Cortese2021}. 
If we divide our sample into central (N=432) and satellites (N=122) according to the classification from the \cite{Yang2007} group catalogue, we find that even in our sample, at fixed q, satellites show a lower $f_{atm}$ (0.2 dex on average) than central galaxies.
This is intriguing, as xGASS includes less extreme density environments than the Virgo cluster analysed in \cite{Li2020}.
However,  even if we focus on central galaxies alone, the scatter in the $f_{atm} - q$ relation remains substantial, with a median 16th to 84th percentile of 0.42 dex.
This means that the large spread in $f_{atm}$ at fixed q does not only trace environmentally-driven gas stripping. 

To further understand the physical origin for this large scatter, we tested many potential galaxy parameters that could be correlated with the vertical offset from the theoretical $f_{atm}$ - q relation, defined as follows;
\begin{equation}
    \Delta f_{atm} = \log_{10}(f_{atm}) - \log_{10}(\rm{min}\{ 1, 2.5 \ q^{1.12} \}).
\end{equation}
Namely, we investigated: the bulge-to-total mass ratio (B/T), Sersic index, \HI\ depletion time ($M_{HI}$/SFR), baryonic mass ($M_{\rm{bar,D}}$), specific SFR and the offset of galaxies from the star-forming main sequence ($\Delta$MS, see Section \ref{section: sample}).
These parameters were chosen as they either related to morphology (which has been shown to relate to the scatter of the Fall relation; \citealt{Romanowsky2012, Cortese2016, Hardwick2022}) or are an aspect of the model that we wanted to make sure was being encapsulated properly.
We use the Spearman rank correlation coefficients ($\rho_{s}$) to quantify the correlation between each quantity and the offset from the theoretical line, as shown in Table \ref{tab: ps} in descending order of correlation. 
We chose to use Spearman rank to measure the correlation as it does not make any assumptions on the type of correlation between the two quantities.
\begin{table}
    \centering
    \begin{tabular}{cc}
                Parameter            & Spearman Correlation Coefficient \\
                \hline
                $\Delta$MS           & $\rho_{s} =  0.43 \pm 0.04$ \\ 
                specific SFR                 & $\rho_{s} =  0.33 \pm 0.04$ \\
                Baryonic Mass        & $\rho_{s} =  0.29 \pm 0.04$ \\  
                \HI\ Depletion time   & $\rho_{s} =  0.25 \pm 0.04$ \\  
                B/T                  & $\rho_{s} = -0.14 \pm 0.04$ \\  
                Sersic index         & $\rho_{s} = -0.11 \pm 0.04$
    \end{tabular}
    \caption{The strength of correlation between $\Delta f_{atm}$ (the vertical offset of our data from the \citetalias{Obreschkow2016} model) and various integrated galaxy properties quantified via the Spearman correlation coefficient and its $1\sigma$ errors (determined from bootstrapping). 
    }
    \label{tab: ps}
\end{table}
The strongest correlation with $\Delta f_{atm}$ is $\Delta$MS, which is known to be strongly linked to a galaxy's atomic gas reservoir \citep[e.g.][]{Catinella2018,Janowiecki2020,Saintonge2022}.
There is also a non-negligible correlation between $\Delta f_{atm}$ and $M_{\rm{bar,D}}$.
This implies that the $f_{atm} - q$ plane may not provide the best representation for the link between AM and gas content in our data, and that xGASS galaxies may be better described by a model with different dependencies on $f_{atm}$ and $M_{\rm{bar,D}}$.
Therefore, the next step is to use a more empirical approach and look for the best way parameterise the correlation between $j_{\rm{bar},D}$, $M_{\rm{bar},D}$ and $f_{atm}$.

\subsection{Empirical 3D fit to j - M - f$_{\rm{atm}}$}

By fitting a plane to specific AM, mass and gas fraction, we aim to quantify what functional form better describes our data and how this differs from the \citetalias{Obreschkow2016} model.
We chose to fit two planes; one for the baryonic parameters ($j_{\rm{bar,D}}$, $M_{\rm{bar,D}}$ and $f_{atm}$) and one for stellar parameters  ($j_{\star,D}$, $M_{\star,D}$ and $f_{\star,atm} = 1.35 M_{HI} / M_{\star,D}$).
Although the majority of this work is focused on baryonic parameters, we also chose to investigate this parameter space using just stellar properties as, a) the stellar parameters have been more extensively studied in the literature \citep[e.g.,][]{Pina2021b,Hardwick2022} and b) they allow us to check if any of our conclusions are affected by the assumptions made to estimate the baryonic AM.
We fit these 3-dimensional planes using the following formula;
\begin{equation}
    \log_{10}(j_{i} / \, \rm{kpc} \,\rm{km}\,\rm{s}^{-1}) = \alpha \log_{10}(M_{i} / \Msol) + \beta \log_{10}(f_{i}) + \gamma
    \label{eq: 3DFit_equation}
\end{equation}
where $i$ is either disc baryons or stars ($\star$).
The best-fitting parameters to equation \ref{eq: 3DFit_equation} are shown in Table \ref{tab: 3Dfits}.
\begin{table*}
    \centering
    \begin{tabular}{cccccc}
                & &     $\alpha$    &    $\beta$      &     $\gamma$  &     $\sigma$ \\
        \hline
        This work & stellar  & $0.81 \pm 0.02$ & $0.38 \pm 0.02$ &  $-5.06 \pm 0.21$ & $0.21 \pm 0.01$ \\
        & baryonic & $0.80 \pm 0.02$ & $0.48 \pm 0.02$ & $-4.86 \pm 0.16$ & $0.15 \pm 0.01$ \\
        \hline
        \citetalias{Obreschkow2016} model & baryonic & 1 & 0.89 & -6.72 &
    \end{tabular}
    \caption{Best-fitting parameters (from equation \ref{eq: 3DFit_equation}) for empirical fits (j - M - gas fraction) to either the stellar component (top row) or the baryonic component (middle row). For comparison, the bottom rows shows the equivalent coefficients for the \citetalias{Obreschkow2016} model.}
    \label{tab: 3Dfits}
\end{table*}
For comparison, when the \citetalias{Obreschkow2016} relation is converted into this formalism, the free parameters of equation \ref{eq: 3DFit_equation} would be $\alpha = 1$, $\beta = 1/1.12 \approx 0.89$ and $\gamma = \log_{10}(G/\, [\rm{kpc}\, ($\kms$)^2\, \Msol^{-1}]) - (1/1.12) \log_{10}(2.5) - \log_{10}(\sigma / [$\kms$]) \approx -6.72$ .
Table \ref{tab: 3Dfits} shows that, as hinted in the previous section, the best-fit  parameters to our data are significantly different from those predicted by the \citetalias{Obreschkow2016} model. 
In particular, there is a slightly weaker dependence on the mass ($\alpha$) term, and a significantly weaker dependence on the gas fraction ($\beta$) term, which is approximately half that expected by the \citetalias{Obreschkow2016} model.
The empirical fit also has a smaller scatter (0.15 dex vertical scatter in $j_{\rm{bar}}$, compared to 0.23 dex vertical scatter observed for our data in the \citetalias{Obreschkow2016} projection).

We obtain very similar results to that of \cite{Pina2021b} who, as mentioned in Section \ref{section: intro}, also fit a 3D plane to baryonic AM, mass and atomic gas fraction. 
The coefficients of our planes are consistent within 3$\sigma$, despite their work focusing on a small sample (N = 157) of gas-rich disc galaxies.
The only major difference between the results of this work and that of \cite{Pina2021b} was that they are very confident in their error analysis and quoted intrinsic scatter of their plane (which did not include measurement errors).
Conversely, as we believe that we are not able to estimate proper errors for each individual galaxy in our sample, we chose to give $\sigma$ which includes both measurement and intrinsic scatter of the plane. 
If we assume that our error estimates are correct, we find enticing evidence for nearly zero intrinsic scatter in our plane, i.e., a plane much tighter than what found in \cite{Pina2021b}. 
However, as mentioned above, these errors are only approximations and we advice caution in over-interpreting this result.

\begin{figure*}
    \centering
    \includegraphics[width=\textwidth]{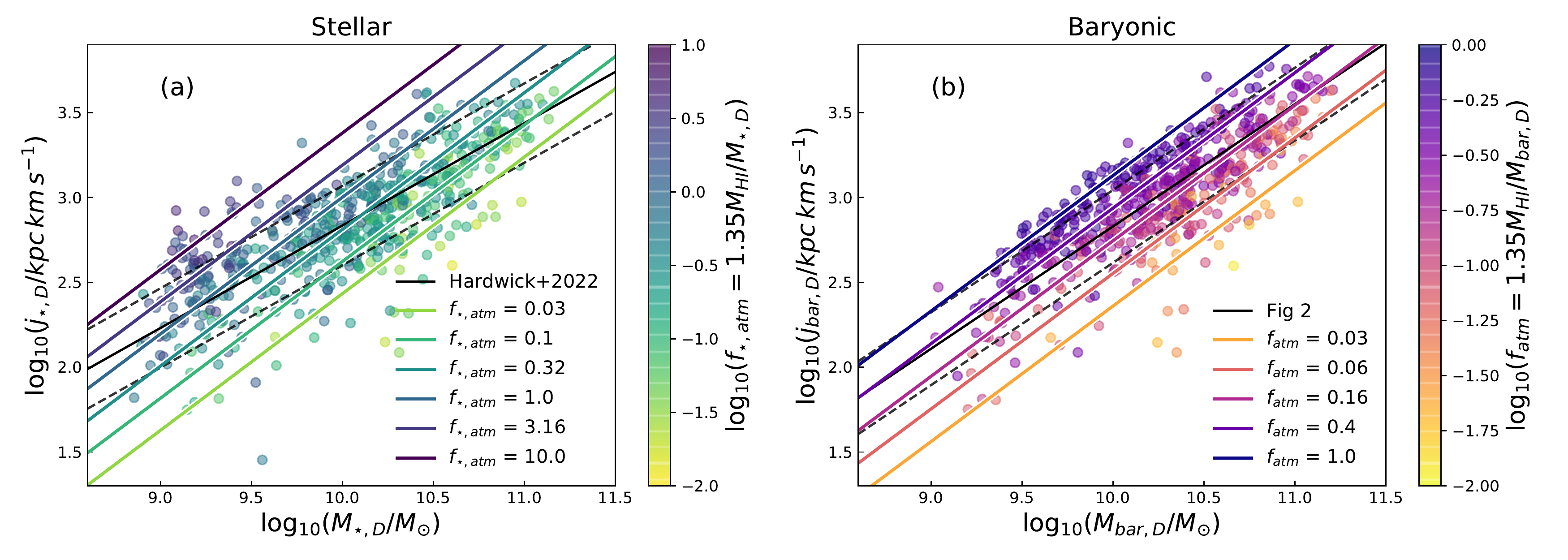}
    \caption{The stellar (a) and baryonic (b) 3D j-M-f$_{atm}$ best-fit relations, in the j - M plane. The planes for fixed gas fractions are shown by the coloured lines.
    Our data points are also colour coded by the gas fraction to compare easily to the best-fits. The best-fitting 2D j-M relations are also shown as thin black lines (with dashed lines showing the 1$\sigma$ vertical scatter).}
    \label{fig: 3DBestFit}
\end{figure*}
Figure \ref{fig: 3DBestFit} shows the stellar (panel a) and baryonic (panel b) relations in the mass - AM plane.
Projections of the best-fitting 3D relations from Table \ref{tab: 3Dfits} are shown for fixed gas fractions as coloured lines. 
To easily compare, points are colour coded in the same way (i.e., by their gas fractions).
The 2D best-fit Fall relation is shown as a thin black line (with the vertical 1$\sigma$ scatter as black dashed lines).
\begin{figure}
    \centering
    \includegraphics[width=0.5\textwidth]{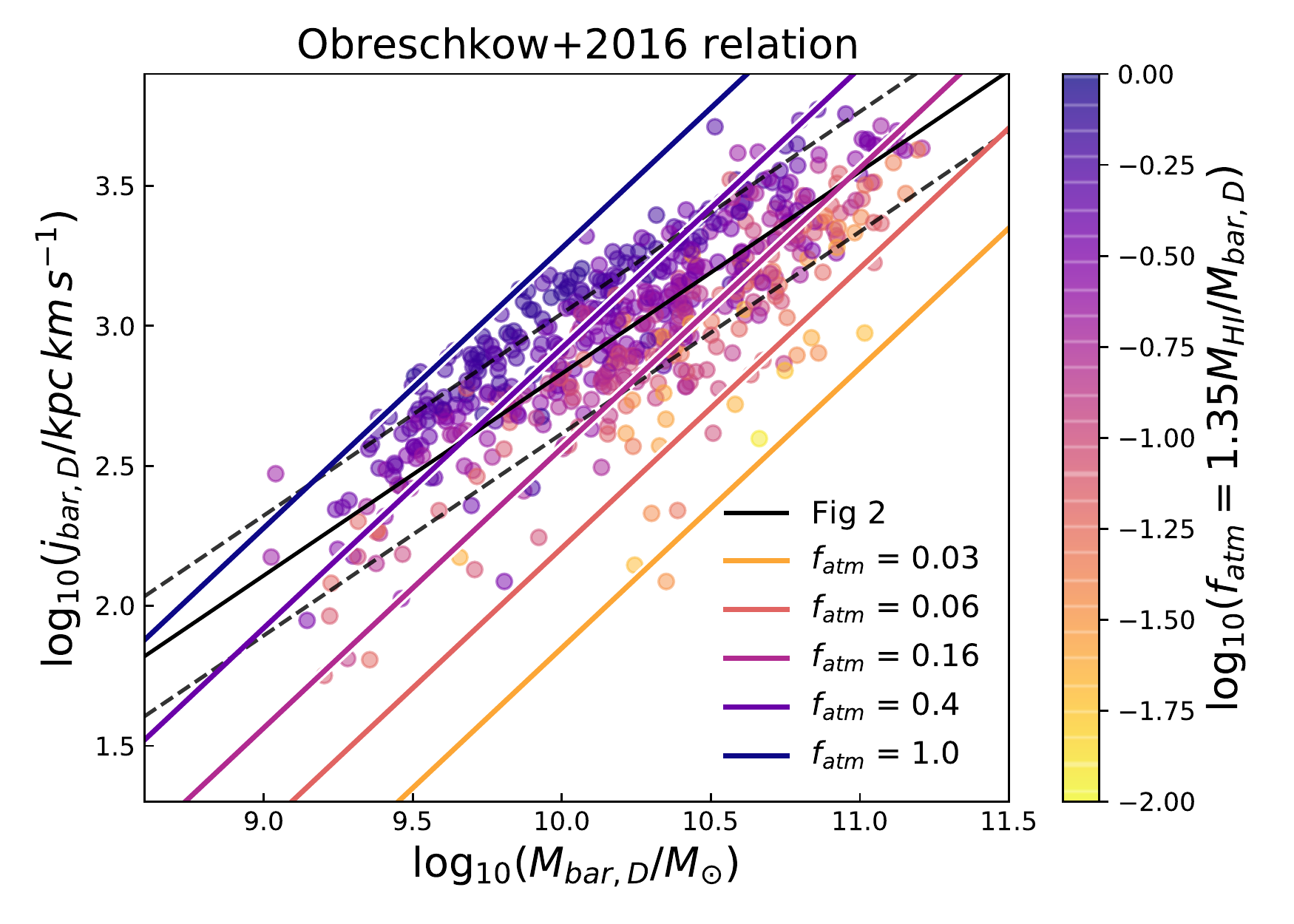}
    \caption{Same as Figure \ref{fig: 3DBestFit}b, but now the coloured lines are projections from the \citetalias{Obreschkow2016} relation.}
    \label{fig: O16Projection}
\end{figure}
For comparison, Figure \ref{fig: O16Projection} shows the same baryonic Fall relation presented in \ref{fig: 3DBestFit}b, but this time the lines of constant gas fraction correspond to the prediction from the \citetalias{Obreschkow2016} model.
The comparison between Figure \ref{fig: 3DBestFit}b and \ref{fig: O16Projection} helps visualising the differences between the \citetalias{Obreschkow2016} model and the best-fit empirical relation.
Specifically, as already mentioned, the empirical relation has a weaker dependence on atomic gas fraction,which can be seen by the larger separation of the lines of constant gas fraction in Figure~\ref{fig: O16Projection}.
The weaker dependence on baryonic mass is also evident, as the lines at fixed gas fraction have a shallower slope than the \citetalias{Obreschkow2016} projection.

A possible issue with the results presented so far is that xGASS was selected to have a nearly flat stellar distribution \citep{Catinella2018, Hardwick2022}, and the oversampling of high mass galaxies may affect our best-fitting results. 
To test this, we also fit the empirical relation including weights for each galaxy. 
Following \cite{Catinella2018}, these weights were calculated to recover the $M_{\star}$ distribution of a volume-limited sample using the local stellar mass function as defined in \cite{Baldry2012}.
We find that, even after including the weights, our best-fit parameters are within error of the values presented in Table~\ref{tab: 3Dfits}.
Therefore, for simplicity, we only present the values obtained without the weights.


Interestingly, the scatter of the empirical relation still shows a weak correlation with $\Delta$MS, with a Spearman rank correlation coefficient of $\rho_{s} = 0.25 \pm 0.05$.
Thus, it is worth exploring whether such residual dependence needs to be included in our fit.
We trialled fitting a 4D plane, with $\Delta$MS as a fourth parameter and found that the dependence on $f_{atm}$ and $M_{\rm{bar},D}$ was largely unchanged when compared to the 3D plane, with a very small dependence ($-0.08 \pm 0.02$) given to the $\Delta$MS term. 
In addition to this, the standard deviation between the 4D and 3D fits was unchanged.
We conclude that any third-order dependence on $\Delta$MS is negligible or at least within the scatter of the 3D fit presented here.

\begin{figure}
    \centering
    \includegraphics[width=0.48\textwidth]{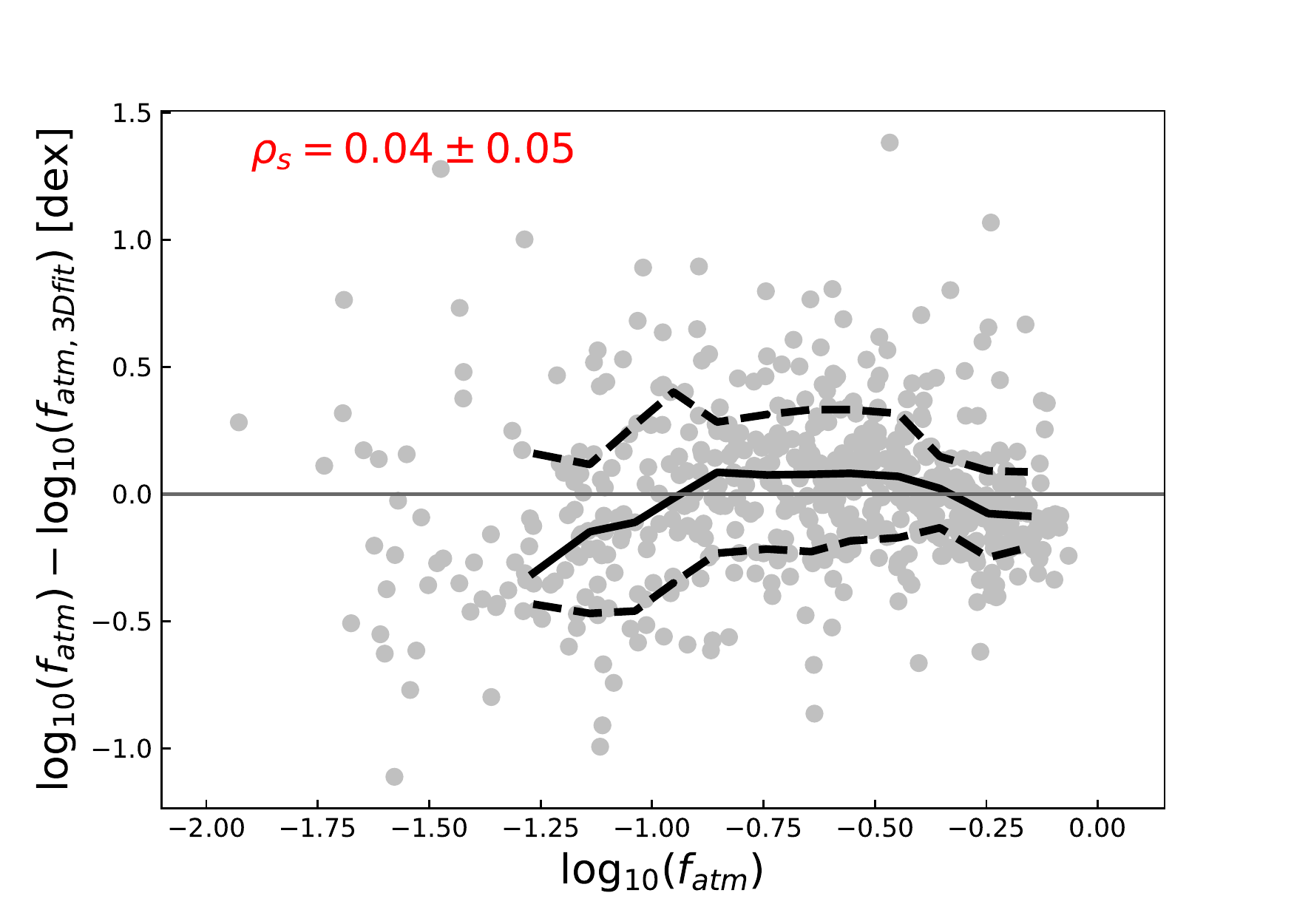}
    \caption{Vertical offset between the $f_{atm}$ measured and that predicted by the empirical plane against  $f_{atm}$. The black thick line shows the median in bins of $f_{atm}$, with each bin having a width of 0.1 dex and bins containing less than 10 galaxies not shown. Dashed lines show 16th and 84th percentiles of those bins.}
    \label{fig: 3D_fatm_offset}
\end{figure}
The choice to use a planar fit for our empirical relation was motivated by the aim of keeping a similar approach to the one adopted by \citetalias{Obreschkow2016}, but we also tested for non-linear trends in the residuals of the empirical fit, to see if this choice was sensible.
There is a slight curvature in the residuals, when binned in $f_{atm}$, as is shown in Figure \ref{fig: 3D_fatm_offset}.
This is a potential indication that the 3D planar fit may be too restrictive. 
However, the curvature is very weak and would likely not change the fit considerably if fit with a curved surface. 
Therefore, while we cannot exclude a curvature in the plane, our sample size is too small to justify the use of a 3D curved surface to parametrise the $j_{\rm{bar}}$, $M_{\rm{bar}}$, $f_{atm}$ plane.

\section{Discussion} \label{section: discussion}

\begin{figure*}
    \centering
    \includegraphics[width=\textwidth]{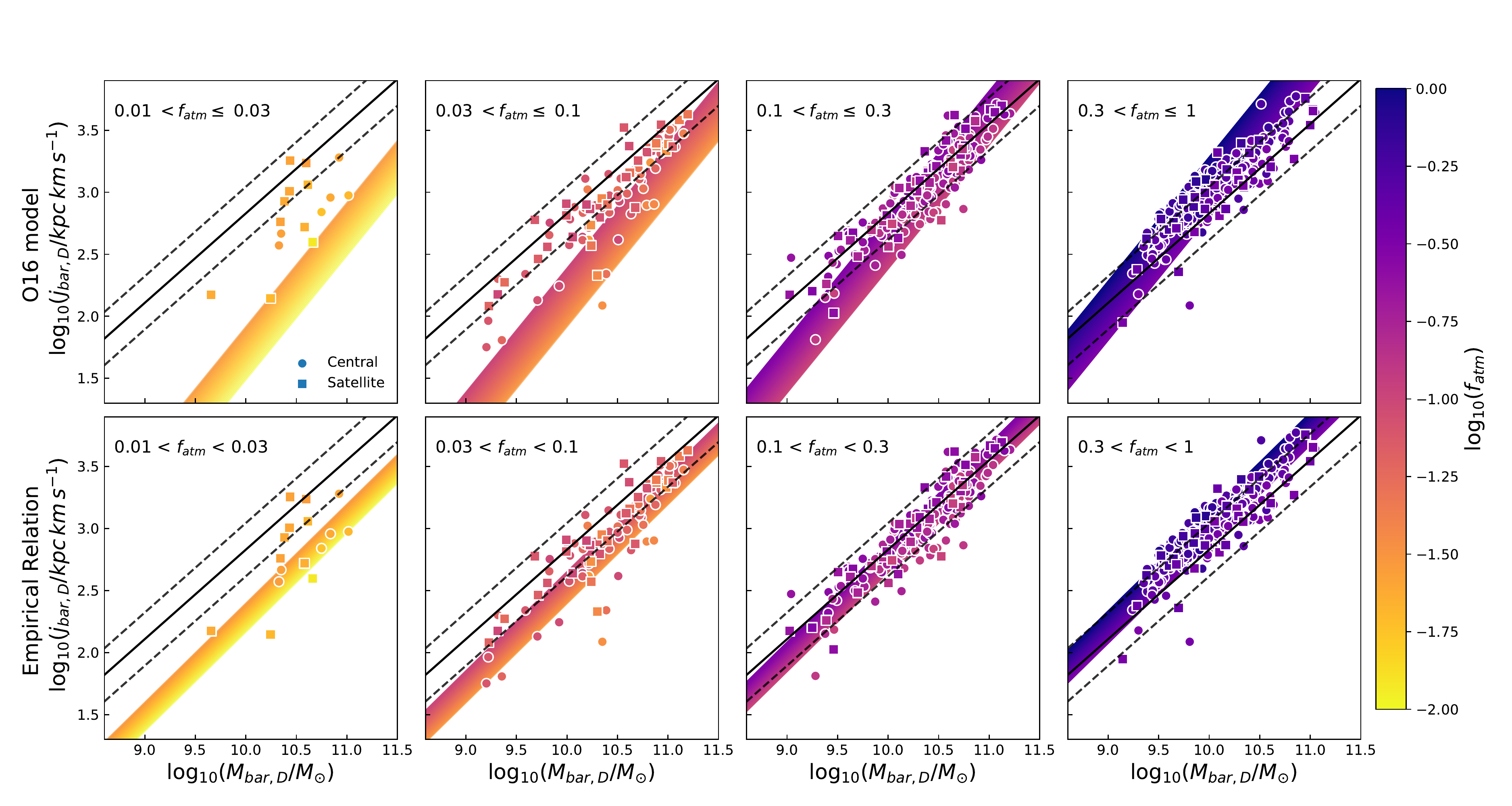}
    \caption{Comparison between the \citetalias{Obreschkow2016} model (top row) and our best-fitting empirical relation (bottom row) in bins of gas fraction. Points are the same in both rows, showing our centrals (circles) and satellite galaxies (squares) colour-coded by their atomic gas fraction. Each column is an even log-spaced bin in gas fraction. The points are comparable to the coloured regions, which show the prediction from either the \citetalias{Obreschkow2016} model or empirical relation for the given gas fraction bin. Overlaid in black for comparison are the best fitting 2D $j_{bar}$-$M_{bar}$ relation from Figure \ref{fig: BaryonicFall}.} 
    \label{fig: Baryonic_fatmbins}
\end{figure*}

Our work shows that the \citetalias{Obreschkow2016} model provides a good first-order representation for the relation between gas fraction and AM. 
However, thanks to the larger dynamic range covered by xGASS compared to previous samples, we have been able to highlight where this analytical model breaks down (in a similar way to \citealt{Pina2021b}).
Specifically, we show that the predicted slope of the baryonic mass - AM relation  at fixed gas fraction is shallower than what is predicted by \citetalias{Obreschkow2016}, with a significantly weaker dependence on gas fraction.  
This is most evident in Figure \ref{fig: Baryonic_fatmbins}, where we directly compare the \citetalias{Obreschkow2016} model and our empirical best-fit  relation to our data at fixed gas fraction. 
In the top row, our data (coloured points) follow a shallower relation than what is predicted by \citetalias{Obreschkow2016} (shaded region), with the largest deviation at lower gas fractions. 

It is important to note that our findings are not a by-product of the methodology adopted to estimate $j_{\rm{bar}}$.
In fact, our main conclusions remain the same if we use stellar quantities alone (see left panel of Figure \ref{fig: 3DBestFit}), showing that our results are not driven by assumptions made on $j_{HI}$ or $j_{\rm{mol}}$. 
Furthermore, the difference between the \citetalias{Obreschkow2016} model and our best-fit  plane is also evident if we examine galaxies in different gas fraction bins separately (as shown in Figure \ref{fig: Baryonic_fatmbins}), for which any systematic variation in the shape of the \HI\ surface density profile should be further reduced \citep[see][]{Wang2016}.

The inability of the \citetalias{Obreschkow2016} model to fully reproduce our data should not come as a surprise. 
This analytic model makes some simplistic assumptions, such as assuming atomic gas and stars are distinct spatially and galaxies are thin axially symmetric equilibrium discs with perfectly circular orbits, any of which could be causing the discrepancies between the model and our sample. 

A similar analytical model has been recently put forward by \cite{Romeo2020}, which uses stability arguments to develop scaling relations involving AM. 
Their model treats each galaxy component (stars, atomic and molecular gas) separately, with each component regulating its own stability, leading to the expression $j_{i} \sigma_{i} / (G M_{i}) \approx 1$ (where $i$ denotes either stars, atomic or molecular gas), which is then used to investigate the scaling relation between AM and gas or stellar mass fraction.
We tested the prediction of the \cite{Romeo2020} model on our sample focusing on atomic gas fraction, and found that our data disagree with the model.
As with the disagreement with \citetalias{Obreschkow2016}, our data prefer a model with a weaker dependence on gas fraction than the \citet{Romeo2020} model predicts.
This is also consistent with the recent work of \cite{Pina2021b}, who found that the \cite{Romeo2020} model was not able to reproduce their data either.

We can therefore go back and focus on the \citetalias{Obreschkow2016} model.
Before proposing any new changes, it is first important to remember that, as discussed in Section \ref{section: fatm-q}, the \citetalias{Obreschkow2016} model does not predict a best-fitting scaling relation, but more simply identifies the parameter space in the $f_{atm}-q$ plane in which galaxies with stable \HI\ discs should exist. 
Galaxies that have too much \HI\ to be stable (i.e., lying above the model) will start rapidly forming stars causing the galaxy to drop down onto the model. 
No such self-regulation is expected below the model, where there is less \HI\ than can be supported by the stability of the disc. 
This \HI\ depletion can be easily achieved by many processes such as environmental processes, outflows, etc. 
Therefore, galaxies are not necessarily expected to be normally distributed around this model and instead preferentially scatter below the relation.
As such a best-fit  to a sample of galaxies that obeys the \citetalias{Obreschkow2016} model will most likely be offset below the \citetalias{Obreschkow2016} model.
If we assume that processes that cause galaxies to scatter below the relation affect all masses equally, then this best-fit  would have the same dependence on $M_{\rm{bar}}$ and $f_{atm}$ as the \citetalias{Obreschkow2016} model.
However, as shown clearly in Figure \ref{fig: 3DBestFit} and \ref{fig: Baryonic_fatmbins}, the best-fit empirical model has a different slope, implying a weaker dependence on baryonic mass and gas fraction than predicted. 
We briefly discuss if any processes could cause these changes. 

One key assumption made in the \citetalias{Obreschkow2016} model is that galaxies live in isolation.
As discussed in Section \ref{section: fatm-q}, environment plays a role in the scatter around the \citetalias{Obreschkow2016} relation, and can also introduce a mass dependence due to environmental processes being more efficient in low mass galaxies.
However, this is unlikely to be the main driver of the deviations, as xGASS does not include galaxies in large cluster environments with the majority of the galaxies in our sample being central/ field galaxies. 
Even when we do exclude satellite galaxies from our sample, we still see a disparity between these galaxies and the \citetalias{Obreschkow2016} model.
Major mergers are also unlikely to be causing this discrepancy, primarily because these are quite rare in the xGASS volume and confused galaxies (including close pairs and merging systems) were excluded from the sample.

The fact that galaxies are not isolated also implies that processes such as smooth gas inflows and outflows (and feedback in general) can play an important role in driving the deviations from the \citetalias{Obreschkow2016} model. 
The effect of these is quite challenging to predict analytically.
However, it is possible that feedback could impact the connection between gas fraction and $q$ by affecting one or more of the key physical quantities at the basis of the \citetalias{Obreschkow2016} model: i.e., gas content, AM and velocity dispersion.

Lastly, the modelling of the 3D structure of galaxies itself most likely harbours some of the limitations of the model. 
Local instabilities caused by spiral arms \citep[e.g.][]{Goldreich1965,Toomre1977,Dobbs2014}, as well as the presence of a significant photometric bulge component in more than half of xGASS galaxies, makes our sample clearly deviating from the simple assumption of axis-symmetric thin discs made in the \citetalias{Obreschkow2016} model.
All this can easily make the assumption of constant \HI\ velocity dispersion incorrect. 
In fact, it has been shown that the \HI\ velocity dispersion radial profiles of nearby galaxies are not constant and that even the average value across the disc varies from galaxy to galaxy \citep[e.g.][]{Bacchini2019}. 
In addition, if the change in $\sigma_{HI}$ correlates with either baryonic mass or gas fraction, it may be even easier to reconcile the tension between our data and the \citetalias{Obreschkow2016} model.
Unfortunately, observational constraints on the variation of $\sigma_{HI}$ with galaxy properties are still missing. 
However, this is an intriguing scenario which could address the issues emerged in this analysis without major changes to the overall physical framework of the \citetalias{Obreschkow2016} model. 

Admittedly, what we discussed so far is primarily speculation, but it is likely that more than one of the issues described above are driving the differences between \citetalias{Obreschkow2016} and our data. 
This is because, as illustrated in Figure \ref{fig: Baryonic_fatmbins}, the tension between our data and the \citetalias{Obreschkow2016} model are present even in the high gas fraction, disc-dominated regime for which this model has been calibrated, and then extends to more gas-poor disc plus bulge systems for which the assumptions of the \citetalias{Obreschkow2016} model naturally break.
This highlights how understanding the physical drivers for the correlation between mass, AM and gas fraction is clearly a complex and multi-dimensional problem. 
The next step towards the solution of this problem may no longer come from a more elaborate analytical model, but instead from investigating this problem in a self-consistent cosmological framework. 
This will be our next step towards fully understanding the connection between AM and gas fraction and the origin of the weaker-than-expected correlation between these properties observed in the xGASS sample.

\section{Summary and Conclusions}
\label{section: summary}

In this work, we have used a representative sample of galaxies in the local Universe with $10^{9} < M_{\star} / \Msol < 10^{11.5}$ to investigate the connection between baryonic mass, specific AM and atomic gas fraction. 
We primarily focus on the comparison of our data with the predictions of the \citetalias{Obreschkow2016} stability model.

We show that, as a first-order approximation, the \citetalias{Obreschkow2016} model agrees well with our sample. 
However, when a full analysis is conducted, we see that our data show a slightly weaker dependence on baryonic mass, and half the dependence on atomic gas fraction than what suggested by the \citetalias{Obreschkow2016} model. 
This implies that the physical connection between AM and gas fraction may be weaker than claimed in previous works, and suggests that some of the assumptions made in the \citetalias{Obreschkow2016} model are not a good representation of the diversity of our data. 
While we speculate that part of this tension may be due to galaxy-to-galaxy variations, it is likely that multiple factors (e.g., internal galaxy properties and environmental effects) are simultaneously responsible for the weaker correlation between AM and gas fraction.
In future work, we plan to further expand this analysis by performing a detailed study of cosmological simulations with the aim of gaining more insights into the exact cause of the discrepancy between our data and the \citetalias{Obreschkow2016} model.

\section*{Acknowledgements}

We thank the anonymous referee for their comments which improved the clarity of our manuscript.
JAH and LC acknowledge support from the Australian Research Council (FT180100066). Parts of this research were conducted by the Australian Research Council Centre of Excellence for All Sky Astrophysics in 3 Dimensions (ASTRO 3D), through project number CE170100013.
DO is a recipient of an Australian Research Council Future Fellowship (FT190100083), funded by the Australian Government.

\section*{Data Availability}

The data used in this work are publicly available at \url{xgass.icrar.org/data.html}.



\bibliographystyle{mnras}
\bibliography{references.bib} 


\bsp	
\label{lastpage}
\end{document}